\documentclass{article}

\usepackage{amsmath}
\usepackage{graphicx,psfrag}
\usepackage[left=2cm,right=2cm,top=2.5cm,bottom=2.5cm]{geometry}
\usepackage{url}

\begin{document}

\title{When Chaos Meets Computers}
\author{Shujun Li\thanks{Reach the author via \texttt{http://www.hooklee.com}.}}
\date{\today}

\maketitle

\begin{abstract}
This paper focuses on an interesting phenomena when chaos meets
computers. It is found that digital computers are absolutely
incapable of showing true long-time dynamics of some chaotic
systems, including the tent map, the Bernoulli shift map and their
analogues, even in a high-precision floating-point arithmetic.
Although the results cannot directly generalized to most chaotic
systems, the risk of using digital computers to numerically study
continuous dynamical systems is shown clearly. As a result, we
reach the old saying that ``it is impossible to do everything with
computers only".
\end{abstract}

\section{Introduction}

This paper focuses on an interesting phenomena when chaos meets
computers. When some chaotic systems are realized in digital
computers, their true long-time dynamics cannot be exhibited at
all, even in a high-precision floating-point arithmetic. Although
the results cannot directly generalized to most other chaotic
systems, the risk of using digital computers to numerically study
continuous dynamical systems is exposed. Actually, in \cite[Chap.
3]{ShujunLi:Dissertation2003}, we have shown some more subtle
dynamical degradation of generic piecewise linear chaotic maps
realized with fixed-point finite precision in digital computers.
Although the phenomena discussed in this paper is completely
different from those studied in \cite[Chap.
3]{ShujunLi:Dissertation2003}, the essential reason of both cases
can be attributed to the use of the multiplication factor 2 and
its powers. Why is 2 a magic number? It is because all digital
computers are based on binary arithmetic, in which a
multiplication with $2^i$ is equal to the left bit-shifting
operation $\ll i$, which may eat $i'\leq i$ precision bits in each
digital chaotic iteration if there does not exist a
bit-compensation operation in the chaotic equation.

Two well-known piecewise linear chaotic maps, the Tent map and the
Bernoulli shift map, are studied in this paper. It is rigorously
proved that all chaotic orbits of the two maps will converge to
zero within a limited number of iterations, and that the average
and maximal numbers of iterations to reach such a convergence are
uniquely determined by details of the involved digital arithmetic.
In addition, it is found that the average number is generally much
smaller than the maximal one, when the initial condition $x(0)$
distributes uniformly in the space of all valid floating-point
numbers. The results on the Tent map and the Bernoulli shift map
can be directly extended to other analogue chaotic maps, including
the V-map $f(x)=2|x-0.5|$, the reflected Bernoulli map
$f(x)=1-(2x\bmod 1)$, and the Baker map (considering Bernoulli
shift map is the $x$-transformation of the Baker map)
\cite{Driebe:ChaoticMaps1999}. Note that the case of the Bernoulli
shift map (also called doubling map) has been well-known to chaos
community \cite{Devaney:Chaos1992}, but the case of the Tent map
not yet. In fact, even for the Bernoulli shift map, there does not
exist a quantitatively analysis on the long-time dynamics of the
map evolving in digital computers. We hope this paper can help to
clarify some questions.

\section{Real Numbers in Digital Computers}

As we know, digital computers adopt binary format to represent
numbers. For real numbers, there are two kinds of representation
formats: fixed-point format, and floating-point format. The
fixed-point format is more suitable for integers or real numbers
with a fixed precision, and the floating-point format is suitable
for real numbers with a higher and variable precision.
Accordingly, there are two kinds of digital arithmetic techniques.
Because most simulation softwares of chaotic systems use
floating-point arithmetic, we will not discuss fixed-point
arithmetic in this paper. Actually, for the problem studied in
this paper, the condition under floating-point arithmetic is much
more complicated than that under fixed-point arithmetic, and the
major results can be easily extended to fixed-point arithmetic.

In today's digital world, the floating-point arithmetic has been
standardized by IEEE and ANSI in 1985
\cite{IEEEStandard754:Floating-Point}, and almost all software and
hardware implementations of floating-point arithmetic obey this
standard to represent the real numbers. In the IEEE/ANSI
floating-point standard, two different floating-point formats are
defined: single-precision and double-precision. The
single-precision format uses 32 bits to represent a real number,
and the double-precision format uses 64 bits. To realize a higher
simulation precision, generally double-precision is used for the
study of chaotic systems. Thus, this paper will focus on
double-precision floating-point arithmetic, and briefly call it
floating-point arithmetic. Note that the extension from
double-precision floating-point arithmetic to single-precision
arithmetic is very easy.

Following the IEEE/ANSI floating-point standard, almost all real
numbers are stored in the following \textit{normalized} format:
\begin{equation}
(-1)^{b_{63}}\times(\overbrace{1.b_{51}\cdots
b_0}^{\mbox{mantissa}})_2\times 2^{\overbrace{(b_{62}\cdots
b_{52})_2-1023}^{\mbox{exponent}}}, \label{equation:normalized}
\end{equation}
where $(\cdot)_2$ means a binary number and $(b_{51}\cdots b_0)_2$
is called the \textit{fraction} of the mantissa. There are five
types of special values that are not represented in the above
normalized format: denormalized numbers, $\pm0$, $\pm\infty$,
indeterminate value, NaN (Not a Number), among which $\pm0$ can be
considered as two special denormalized numbers. Note that $+0\neq
-0$ and $+\infty\neq -\infty$. The five types of special values are
stored in the following formats
\cite{IEEEStandard754:Floating-Point,Hollasch:FloatingPointStd}:
\begin{itemize}
\item \textit{denormalized numbers}: when all exponent bits are
zeros, i.e., $b_{62}=\cdots=b_{52}=0$, the floating-point number
does not yield Eq. (\ref{equation:normalized}), but represents the
following fixed-point number:
\begin{equation}
(-1)^{b_{63}}\cdot(0.b_{51}\cdots b_0)_2\times
2^{-1022}.\label{equation:denormalized}
\end{equation}
Apparently, denormalized numbers can be considered as fixed-point
numbers whose absolute values are lower than the minimal positive
normalized number $\left(1\times 2^{1-1023}=2^{-1022}\right)$.

\item $\pm0$: the reason that denormalized numbers are used is the
existence of the hidden 1-bit in the normalized format, which
makes it impossible to represent zero. With the denormalized
format, the representation of zeros becomes natural:
$b_{62}=\cdots=b_0=0$, and $b_{63}$ serves as the sign bit.

\item $\pm\infty$ are represented as follows:
$b_{62}=\cdots=b_{52}=1$, $b_{51}=\cdots=b_0=0$, and $b_{63}$ is
used to denote the sign.

\item \textit{NaN} (Not a Number) can be represented with all other invalid
formats, which include a) QNaN (Quite NaN): $b_{62}=\cdots=b_{52}=1$
and $(0\cdots 01)_2\leq(b_{51}\cdots b_0)_2\leq(01\cdots 1)_2$; b)
SNaN (Signalling NaN): $b_{62}=\cdots=b_{52}=1$ and $(10\cdots
0)_2\leq(b_{51}\cdots b_0)_2\leq(11\cdots 1)_2$.
\end{itemize}

\section{The Studied Chaotic Maps}

\begin{figure}[!htb]
\centering
\begin{minipage}{0.45\textwidth}
\centering
\includegraphics[width=\textwidth]{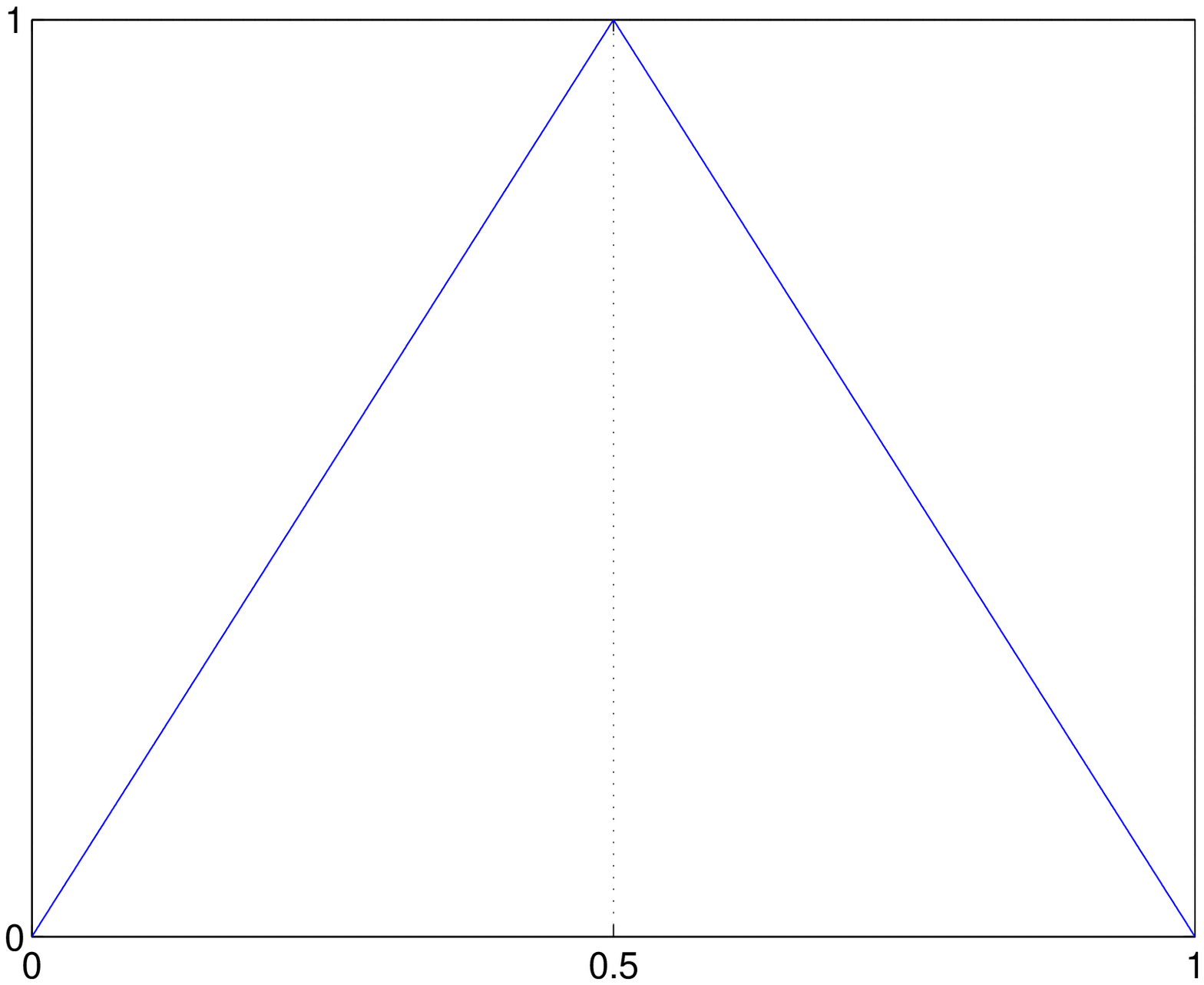}
a) the Tent map
\end{minipage}
\begin{minipage}{0.45\textwidth}
\centering
\includegraphics[width=\textwidth]{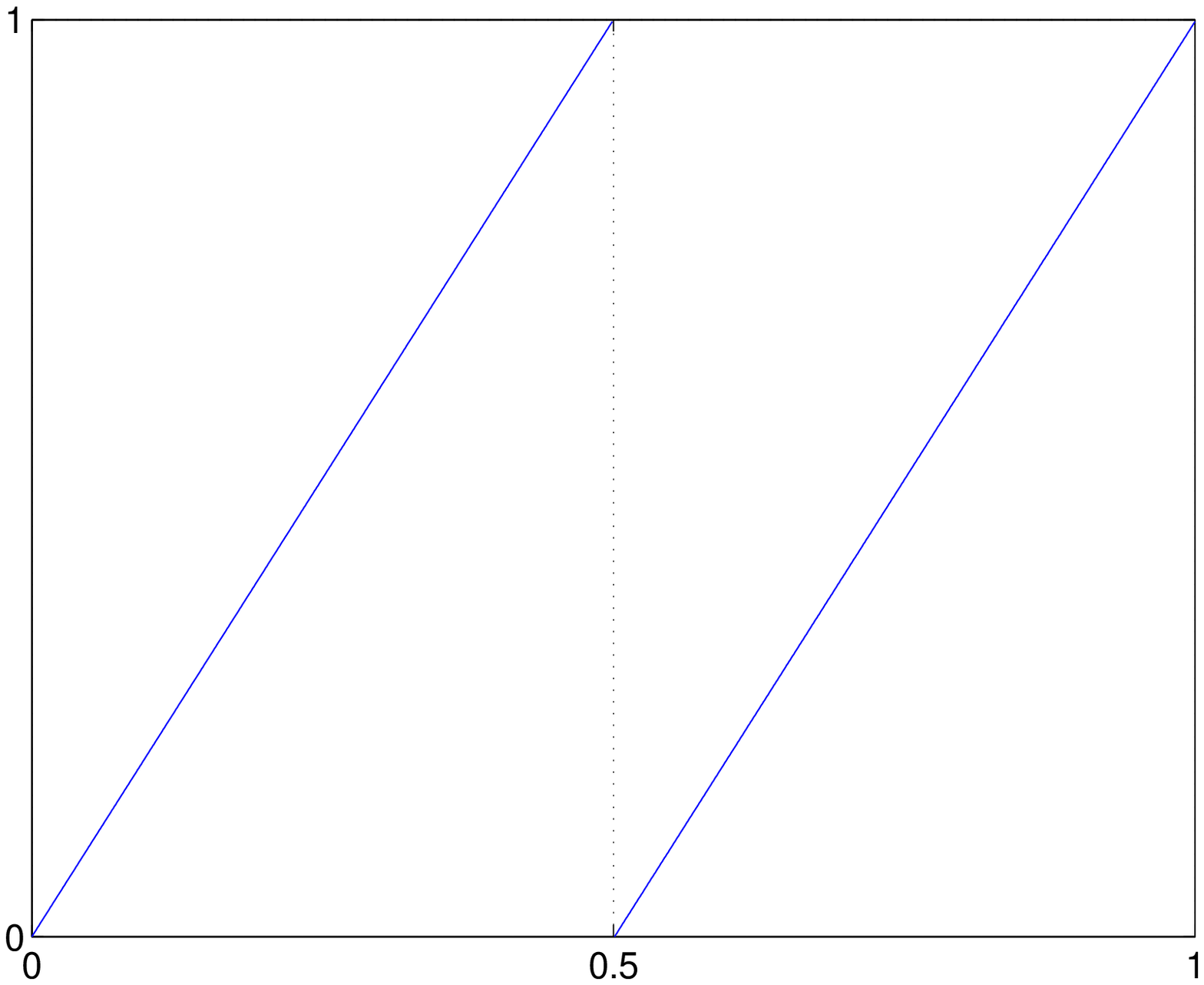}
b) the Bernoulli shift map
\end{minipage}
\caption{The Tent map and the Bernoulli shift
map}\label{figure:Tent}
\end{figure}

In this paper, two well-known discrete-time chaotic maps are
studied to show the incapability of digital computers: the Tent
map and the Bernoulli shift map \cite{Driebe:ChaoticMaps1999}. The
Tent map is defined as
\begin{equation}
f(x)=1-2|x-0.5|=\begin{cases}
2x, & 0\leq x<0.5,\\
2(1-x), & 0.5\leq x\leq 1,
\end{cases}
\end{equation}
and the Bernoulli shift map is defined as
\begin{equation}
f(x)=2x\bmod 1=\begin{cases}
2x, & 0\leq x< 0.5,\\
2x-1, & 0.5\leq x<1,\\
0, & x=1,
\end{cases}
\end{equation}
which are shown in Fig. \ref{figure:Tent}a and \ref{figure:Tent}b,
respectively.

As we know, the two simple chaotic maps have typical chaotic
dynamics \cite{ShujunLi:Dissertation2003,Baranovsky:PLCM:IJBC95}:
1) positive Lyaponuv exponents; 2) ergodicity, mixing and
exactness; 3) uniform invariant density function $f^*(x)=1$; 4)
the Tent map's auto-correlation function $\tau_t(n)=\delta(n)$,
and the Bernoulli shift map's auto-correlation function
$\tau_B(n)$ approaches to zero exponentially as $n\to\infty$. The
third property means that the chaotic orbit starting from almost
everywhere will lead to the same uniform distribution $f^*(x)=1$
over the definition interval [0,1]. However, as shown below, this
paper points out that such a good dynamics cannot be exhibited in
digital computers well, due to the finite-precision effect.

\section{Chaotic Maps under Floating-Point Arithmetic}
\label{section:Analysis}

Firstly, assume the initial condition $x(0)=(0.b_1b_2\cdots
b_j\cdots b_{L-1}b_L)_2\neq 0$, where $b_L=1$ (the least significant
1-bit) and $1-x(0)=(0.b_1'b_2'b_3'\cdots b_j'\cdots b_{L-1}'b_L)_2$.
Then, the iteration of the Tent map will be
\begin{equation}
x(1)=\begin{cases} 2x(0)=x(0)\ll 1=(0.b_2\cdots b_j\cdots b_{L-1}b_L)_2, & 0\leq x(0)<0.5,\\
2(1-x(0))=(b_1'.b_2'b_3'\cdots b_j'\cdots b_{L-1}'b_L)_2, &
0.5\leq x(0)\leq 1,
\end{cases}
\end{equation}
where $\ll$ denotes the left bit-shifting operation. Note that
$b_1=0$ when $0\leq x(0)<0.5$. Apparently, after $L-1$ iterations,
$x(L-1)\equiv(0.b_L)_2=(0.1)_2$. Then, $x(L)\equiv 1$, and
$x(L+1)\equiv 0$. That is, the number of required iterations to
converge to zero is $N_r=L+1$. Note that $N_r=0$ when $x(0)=0$.

For the Bernoulli shift map, we can similarly get
\begin{equation}
x(1)=2x(0)\bmod 1=x(0)\ll1=(0.b_2b_3\cdots b_j\cdots
b_{L-1}b_L)_2.
\end{equation}
Apparently, after $N\geq L$ iterations, $x(N)\equiv0$. That is, the
number of required iterations to converge to zero is $N_r=L$. Note
that $N_r=0$ when $x(0)=0$.

From the above analysis, it is clear that no any quantization
error is introduced in the digital chaotic iterations, which is
because the chaotic iterations can be exactly carried out with the
digital operation $\ll$.

In the following, let us consider the value of $L$ in two
different conditions of $x(0)\neq 0$:
\begin{itemize}
\item
\textit{$x(0)$ is a normalized number}: from Eq.
(\ref{equation:normalized}), $x(0)=(1.b_{51}\cdots b_0)_2\times
2^{-e}=(0.\overbrace{0\cdots 0}^{e-1}1b_{51}\cdots b_0)_2$.
Assuming the least 1-bit of $x(0)$ is $b_i=1$, one can immediately
get $x(0)=(0.\overbrace{0\cdots 0}^{e-1}1\overbrace{b_{51}\cdots
b_i}^{52-i}\overbrace{0\cdots 0}^i)_2$ and deduce
$L=(e-1)+1+(52-i)=e+(52-i)$. Considering $1\leq e\leq 1022$ and
$0\leq i\leq 51$, $2\leq L\leq 1074$.

\item
\textit{$x(0)$ is a non-zero denormalized number}: from Eq.
(\ref{equation:denormalized}), $x(0)=(0.b_{51}\cdots b_0)_2\times
2^{-1022}=(0.\overbrace{0\cdots 0}^{1022}b_{51}\cdots b_0)_2$.
Assuming the least 1-bit of $x(0)$ is $b_i=1$, one can immediately
get $x(0)=(0.\overbrace{0\cdots 0}^{1022}\overbrace{b_{51}\cdots
b_i}^{52-i}\overbrace{0\cdots 0}^i)_2$ and deduce
$L=1022+(52-i)=1074-i$. Considering $0\leq i\leq 51$, $1023\leq
L\leq 1074$.
\end{itemize}
To sum up, in both conditions $L\leq 1074$.

In the following, we will prove that the mathematical expectation
of $L$ is only about much smaller than 1074 if $x(0)$ distributes
uniformly in the space of all valid floating-point numbers in
[0,1]. That is, the mathematical expectation of $N_r$ is much
smaller than 1074 for the Bernoulli shift map, and smaller than
1075 for the Tent map.

Firstly, let us consider the mathematical expectation of
$i\in\{0,\cdots,51\}$. Without loss of generality, for a
denormalized number or a normalized number with a fixed exponent
$e$, assume the mantissa fraction $(b_{51}\cdots b_0)_2$ distributes
uniformly over the discrete set $\{0,\cdots,2^{52}-1\}$. Then, the
probability that $(b_i=1,b_{i-1}=\cdots=b_0=0)$ is
$\dfrac{2^{51-i}}{2^{52}}=\dfrac{1}{2^{i+1}}$, and the probability
that $(b_{51}=\cdots=b_0=0)$ is $\dfrac{1}{2^{52}}$. Then, the
mathematical expectation of $i$ is
\begin{equation}
E(i)\approx\sum_{i=0}^{51}i\cdot\frac{1}{2^{i+1}}+52\cdot\frac{1}{2^{52}}=
\frac{1}{2}\cdot\sum_{i=1}^{51}\frac{i}{2^{i}}+\frac{52}{2^{52}}=
\frac{1}{2}\cdot\left(2-\frac{53}{2^{51}}\right)+\frac{52}{2^{52}}=1-\frac{1}{2^{52}}\approx
1.
\end{equation}

Then, let us consider the mathematical expectation of
$e\in\{1,\cdots,1022\}$. From the uniform distribution of $x(0)$
in the interval [0,1], one has the probability of the exponent is
$e$ is about $Prob[2^{-e}\leq x<2^{-(e-1)}]=2^{-e}$. Thus, the
mathematical expectation of $e$ is
\begin{equation}
E(e)\approx\sum_{e=1}^{1022}\frac{e}{2^e}=2-\frac{1024}{2^{1022}}\approx
2.
\end{equation}

From the above deductions, we can immediately deduce
\begin{eqnarray*}
E(L) & = & Prob[\mbox{normalized numbers}]\cdot(E(e)+(52-E(i)))\\
& & {}+Prob[\mbox{denormalized numbers}]\cdot(1074-E(i))\\
& = & \frac{1022}{1023}\cdot(E(e)+(52-E(i)))+\frac{1}{1023}\cdot(1074-E(i))\\
& \approx & \frac{1022}{1023}\cdot(2+(52-1))+\frac{1}{1023}\cdot(1074-1))\\
& = & \frac{1022}{1023}\cdot 53+\frac{1}{1023}\cdot 1073\\
& = & 53+\frac{1020}{1023}\approx 53.997.
\end{eqnarray*}
Generally denormalized numbers will not be used by most
pseudo-random number generators, such as the embedded
\texttt{rand} function in almost all programming languages, so
$E(L)=E(e)+(52-E(i))\approx 53$. That is, $E(N_r)\approx 53<<
1074$ for the Bernoulli shift map, and $E(N_r)\approx 54<<1075$
for the Tent map, where $a<<b$ means that $a$ is much smaller than
$b$.

\section{Experiments}

\begin{figure}[!htb]
\centering\psfrag{i}{$i$}\psfrag{x(i)}{$x(i)$}\psfrag{1074+1=1075
}{$1074+1=1075$}\psfrag{1022+53=1075 }{$1022+53=1075$}
\includegraphics[width=0.6\textwidth]{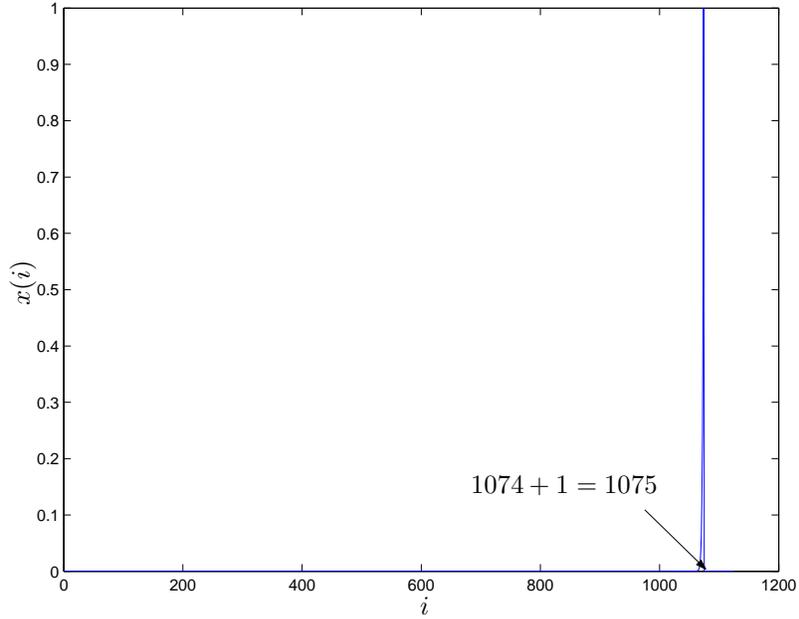}
\caption{The evolution of the digital Tent map with the initial
condition $x(0)=2^{-1074}$}\label{figure:TestNumber1}
\end{figure}

\begin{figure}[!htb]
\centering\psfrag{i}{$i$}\psfrag{x(i)}{$x(i)$}\psfrag{1022+53=1075
}{\hspace{-2em}$1022+53=1075$}
\includegraphics[width=0.6\textwidth]{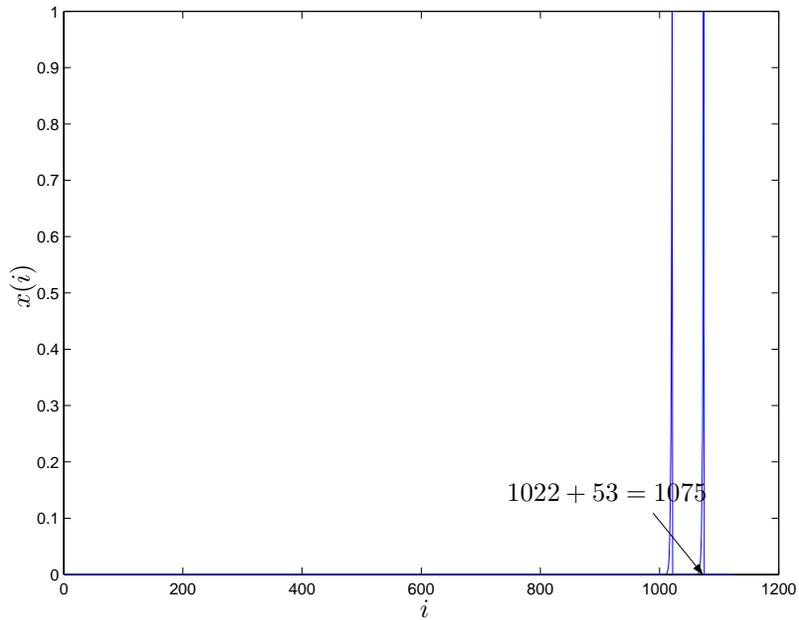}
\caption{The evolution of the digital Tent map with the initial
condition $x(0)=(1.1\cdots 1)_2\times
2^{-1022}$}\label{figure:TestNumber2}
\end{figure}

To verify the analysis given in the last section, a larger number of
floating-point numbers are tested for the two maps. Taking the Tent
map as an example, the results of three typical numbers, the minimal
denormalized number $x(0)=(\overbrace{0\cdots
0}^{63}1)_2=2^{-1074}$, and a normalized number
$x(0)=(0|\overbrace{0\cdots0}^{10}1|\overbrace{1\cdots
1}^{52})_2=(1.\overbrace{1\cdots 1}^{52})_2\times 2^{-1022}$ are
given in Figs. \ref{figure:TestNumber1} and
\ref{figure:TestNumber2}, respectively.

\begin{figure}[!htb]
\centering
\includegraphics[width=0.6\textwidth]{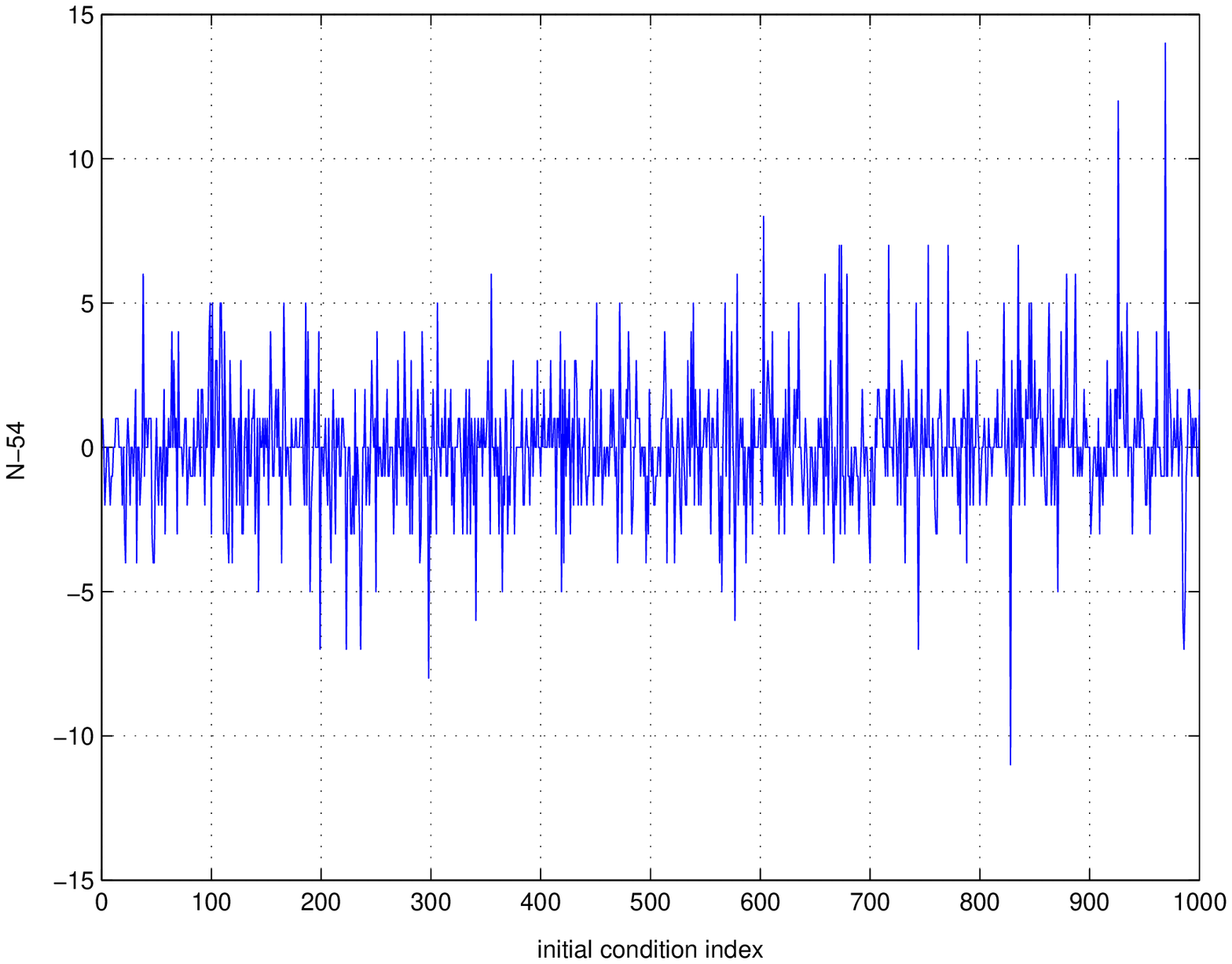}
\caption{The values of $N_r$ for 1000 randomly-generated initial
conditions}\label{figure:Tent1000ICs}
\end{figure}

\begin{figure}[!htb]
\centering
\includegraphics[width=0.6\textwidth]{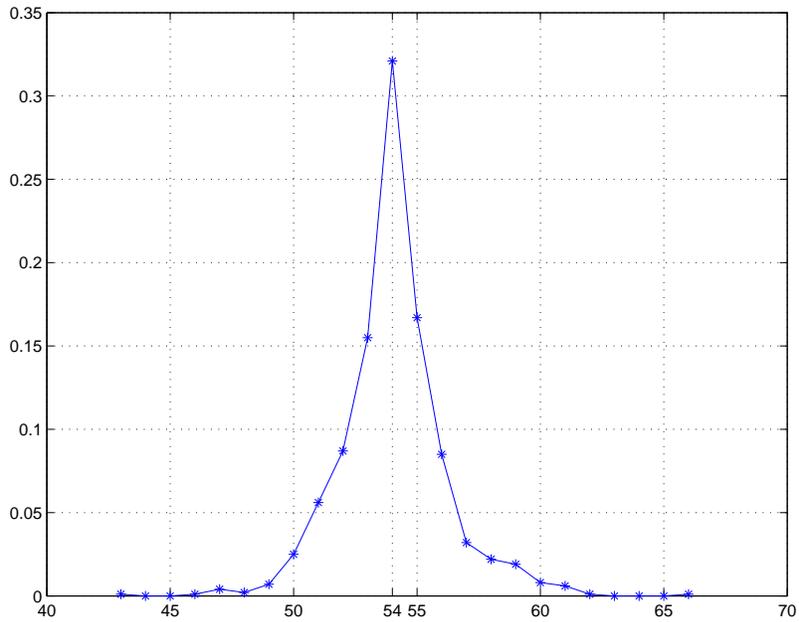}
\caption{The occurrence frequency of different values of $N_r$ in
the total 1000 values}\label{figure:Tent1000ICs-hist}
\end{figure}

To verify the mathematical expectation of $N_r$, some experiments
are made for test on 1000 initial conditions pseudo-randomly
generated with the standard \texttt{rand} function of matlab. The
values of $N_r$ corresponding to the 1000 initial conditions are
shown in Fig. \ref{figure:Tent1000ICs}, and the occurrence frequency
of different values is shown in Fig. \ref{figure:Tent1000ICs-hist}.
Following the data, we can get the average of $N_r$ is about
$54.03\approx E(N_r)$, which agrees with the theoretical result
well.

\section{Conclusions}

This paper studies the digital iterations of the two well-known
chaotic discrete-time maps -- the Tent map and the Bernoulli shift
map. It is found that all chaotic orbits will be eventually
converge to zero within $N_r$ iterations, and that the value of
$N_r$ is uniquely determined by the details of digital
floating-point arithmetic. Although the given results are rather
simple from a computer scientist's point of view, they really
touches the subtle kernel of digital chaos (i.e., chaos in
computers).

\bibliographystyle{unsrt}
\bibliography{test}

\begin{thebibliography}{1}

\bibitem{ShujunLi:Dissertation2003}
Shujun Li.
\newblock {\em Analyses and New Designs of Digital Chaotic Ciphers}.
\newblock PhD thesis, School of Electronic and Information Engineering, Xi'an
  Jiaotong University, Xi'an, China, June 2003.
\newblock available online at \url{http://www.hooklee.com/pub.html}.

\bibitem{Driebe:ChaoticMaps1999}
Dean~J. Driebe.
\newblock {\em Fully Chaotic Maps and Broken Time Symmetry}, volume~4 of {\em
  Nonlinear Phenomena and Complex Systems}.
\newblock Kluwer Academic Publishers, Dordrecht, The Netherlands, 1999.

\bibitem{Devaney:Chaos1992}
Robert~L. Devaney.
\newblock {\em A First Course in Chaotic Dynamical Systems: Theory and
  Experiment}.
\newblock Addison-Wesley Publishing Company, Inc., Reading, Massachusetts,
  1992.

\bibitem{IEEEStandard754:Floating-Point}
IEEE~Computer Society.
\newblock {IEEE} standard for binary floating-point arithmetic.
\newblock ANSI/IEEE Std. 754-1985, August 1985.

\bibitem{Hollasch:FloatingPointStd}
Steve Hollasch.
\newblock {IEEE} standard 754 floating point numbers.
\newblock online document at
  \url{http://stevehollasch.com/cgindex/coding/ieeefloat.html}, February 2005.

\bibitem{Baranovsky:PLCM:IJBC95}
A.~Baranovsky and D.~Daems.
\newblock Design of one-dimensional chaotic maps with prescribed statistical
  properties.
\newblock {\em Int. J. Bifurcation and Chaos}, 5(6):1585--1598, 1995.

\end{thebibliography}

\end{document}